\documentclass{WileyMSP-template}
\usepackage{color}
\usepackage{graphicx}
\usepackage{cite}
\usepackage{amsmath}
\begin{document}

\pagestyle{fancy}
\rhead{\includegraphics[width=2.5cm]{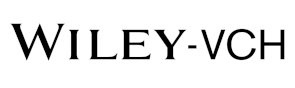}}
\rmfamily

\title{Reconfigurable Terahertz Multi-Harmonic Dual-Combs}

\maketitle


\author{Xuhong Ma*,**}
\author{Zhiwei Qin*}
\author{Xianglong Bi}
\author{Ziping Li}
\author{Wenjian Wan}
\author{Binbin Liu}
\author{Guibin Liu}
\author{Yue Pan}
\author{Yanming Lu}
\author{Hua Li**}


\dedication{$^{*}$ These authors contributed equally to this work.}\\
\dedication{$^{**}$ Corresponding author. Email: maxuhong@mail.sim.ac.cn and hua.li@mail.sim.ac.cn.}

\begin{affiliations}
Dr. X. Ma, Z. Qin, X. Bi, Dr. Z. Li, Dr. W. Wan, Dr. B. Liu, G. Liu, Y. Pan, Y. Lu, Prof. H. Li\\
National Key Laboratory of Materials for Integrated Circuits and Key Laboratory of Terahertz Solid State Technology, Shanghai Institute of Microsystem and Information Technology, Chinese Academy of Sciences, 865 Changning Road, Shanghai 200050, China\\
   E-mail: maxuhong@mail.sim.ac.cn, hua.li@mail.sim.ac.cn.

Z. Qin, X. Bi, Dr. Z. Li, Dr. W. Wan, G. Liu, Y. Pan, Y. Lu, Prof. H. Li\\
Center of Materials Science and Optoelectronics Engineering, University of Chinese Academy of Sciences, Beijing 100049, China

Dr. X. Ma, Dr. B. Liu\\
Chongqing Key Laboratory of Precision Optics, Chongqing
Institute of East China Normal University, Chongqing 401120,
China

\end{affiliations}



\begin{abstract}

Dual-comb spectroscopy, constructed from two frequency combs with slightly different repetition frequencies, enables real-time and high-precision measurements without mechanical scanning. In the terahertz (THz) spectral range, dual-comb techniques provide a powerful tool for high-resolution and rapid spectroscopy. Quantum cascade lasers (QCLs), owing to their compact footprint and favorable size, weight, and power (SWaP), have emerged as promising sources for THz dual-comb systems. In QCL frequency combs, the repetition frequency generated through intrinsic four-wave mixing is typically equal to the cavity round-trip frequency, corresponding to the fundamental comb. Although this repetition frequency can be tuned via current and temperature control, the accessible tuning range remains limited. Recently, harmonic frequency combs, whose repetition frequencies are integer multiples of the cavity round-trip frequency, have attracted increasing attention, offering enhanced single-line signal-to-noise ratios and providing direct insight into the strong nonlinearity of QCLs. Here, we demonstrate a reconfigurable multi-harmonic dual-comb system realized on a single self-detected THz QCL platform. By precisely controlling the driving current and thermal conditions, we achieve and switch between multiple dual-comb configurations, including fundamental–fundamental, fundamental–second-harmonic, second-harmonic–second-harmonic, and second-harmonic–third-harmonic dual-combs. These results establish harmonic order as an additional degree of freedom for dual-comb operation within a single laser system. The demonstrated platform enables high-resolution and high-sensitivity measurements while significantly simplifying the system architecture. More importantly, it reveals the pronounced intrinsic nonlinearity of THz QCLs, allowing dual-comb generation even with weak spectral overlap. This reconfigurable multi-band dual-comb approach based on a single QCL opens new opportunities for compact THz spectroscopy and frequency metrology.

\end{abstract}


\section{Introduction}

Quantum cascade lasers (QCLs) operating in the mid-infrared and terahertz (THz) spectral regions have emerged as compact and powerful optical frequency comb sources\cite{hugi2012mid,rosch2015octave,rosch2018heterogeneous,faist2016quantum}. Benefiting from their small footprint, low power consumption, and strong intrinsic nonlinearity\cite{faist2016quantum,williams2007terahertz,scalari2009thz,vitiello2015quantum}, QCLs can sustain self-mode-locked comb operation in a free-running regime without external modulation\cite{hugi2012mid,faist2016quantum,burghoff2015evaluating,burghoff2014terahertz,khurgin2014coherent}. In particular, Fabry–Pérot (FP) cavity QCLs support frequency comb formation through spontaneous four-wave mixing, which phase-locks longitudinal modes and produces evenly spaced spectral lines with repetition frequencies equal to the cavity free spectral range (FSR, the frequency spacing between adjacent longitudinal modes). Since the FSR can be tuned via electro-thermal control of the refractive index and cavity length, QCL combs provide a compact and electrically controllable platform for precision spectroscopy and frequency metrology\cite{villares2015chip,hillbrand2019coherent,hayden2024mid}.

Dual-comb spectroscopy, based on two frequency combs with slightly detuned repetition frequencies, enables rapid and high-resolution measurements without mechanical scanning\cite{villares2014dual,coddington2010coherent,schliesser2012mid,picque2019frequency}. In the THz regime, where fast and sensitive detectors remain technologically challenging, QCL-based dual-comb systems offer a distinct advantage\cite{williams2007terahertz,scalari2009thz,vitiello2015quantum,barbieri20042,liu2025terahertz}. Owing to the ultrafast carrier dynamics (picosecond-scale gain recovery time), a QCL can simultaneously act as both an emitter and a detector via the self-detection mechanism, enabling a highly simplified dual-comb configuration using two FP QCLs\cite{villares2014dual,rosch2016chip,yang2016terahertz,consolino2019fully}. This architecture has enabled MHz-level frequency resolution and stimulated extensive developments in stabilization\cite{consolino2019fully,liao2022broadband,liu2025farey}, phase correction\cite{li2023terahertz,coddington2010coherent}, cavity-length tuning\cite{zhao2021active,villares2015chip}, and system integration\cite{wang2021improved,villares2014dual}, significantly advancing practical THz dual-comb applications\cite{senica2022planarized,wu2026mutual}.

Nevertheless, despite these advantages, current THz QCL dual-comb systems still suffer from limited flexibility in spectral and frequency-domain control. Within a finite gain bandwidth, increasing the number of comb lines inherently reduces the optical power per line, leading to a trade-off between spectral coverage and signal-to-noise ratio (SNR)\cite{hayden2024mid,komagata2023absolute,faist2016quantum}. This constraint becomes particularly critical in applications requiring high sensitivity or large dynamic range, where weak absorption features or low-responsivity states must be resolved.

This limited controllability is further exacerbated by the lack of mature THz optical components and spectral manipulation techniques\cite{tan2026continuous}. The GHz-level mode spacing of THz QCL combs remains too small for mature THz narrowband filtering and line-by-line spectral manipulation, making it difficult to selectively address individual comb lines with available optical components. As a result, advanced functionalities such as frequency-to-space mapping, parallel spectral processing, or multi-channel coherent communication remain difficult to implement in the THz domain\cite{dhillon20172017,lewis2014review,tonouchi2007cutting}. This limitation highlights the need for alternative approaches to spectral control and mapping that do not rely on external optical components.

To overcome these limitations, recent investigations into the nonlinear dynamics of QCL comb formation have revealed the emergence of harmonic frequency combs. In these states, the repetition frequency becomes an integer multiple of the cavity FSR, resulting in fewer but more widely spaced comb lines\cite{piccardo2018widely,silvestri2022multimode,justo2024harmonic}. The emergence of harmonic combs in QCLs is attributed to the interplay between gain competition, spatial hole burning, and the unique gain recovery dynamics in intersubband transitions\cite{2021APLselfstarting,LiOE2022}. Recently, Silvestri et al. have further revealed that such harmonic states can emerge from a resonance between the effective Rabi frequency and a cavity mode of the QCL, providing a deeper theoretical framework for understanding the spontaneous formation of harmonic frequency combs in ultrafast semiconductor lasers\cite{carlo2026Rabi}. Although the detailed physical mechanisms remain under active study, harmonic combs effectively modify the spectral distribution within the same gain bandwidth and provide an alternative route for repetition-frequency engineering\cite{wang2020harmonic}. From a dual-comb perspective, the harmonic order introduces a new degree of freedom for repetition-frequency engineering, providing an alternative route to frequency mapping and spectral control without relying on external optical components.

Building on this concept, in this work we demonstrate a reconfigurable multi-harmonic dual-comb system realized within a single self-detected THz QCL platform. By precisely controlling the driving current and thermal conditions, we selectively generate and switch among multiple dual-comb configurations, including fundamental–fundamental, fundamental–second-harmonic, second-harmonic–second-harmonic, and second-harmonic–third-harmonic dual-combs. This approach establishes harmonic order as an additional degree of freedom for dual-comb operation, enabling flexible and reconfigurable frequency down-conversion schemes without duplicating sources or relying on external optical components.

The demonstrated architecture highlights the strong intrinsic nonlinearity of THz QCLs, enabling dual-comb operation even with weak spectral overlap between harmonic states. By consolidating multi-regime dual-comb functionality within a single laser device, this work provides a compact and versatile platform for THz spectroscopy and frequency metrology, with enhanced flexibility for future system-level integration and spectral control.

\section{Results and Discussion}

\subsection{Experimental setup}

\begin{figure}[!h]
 \centering
 \includegraphics[width=0.9\linewidth]{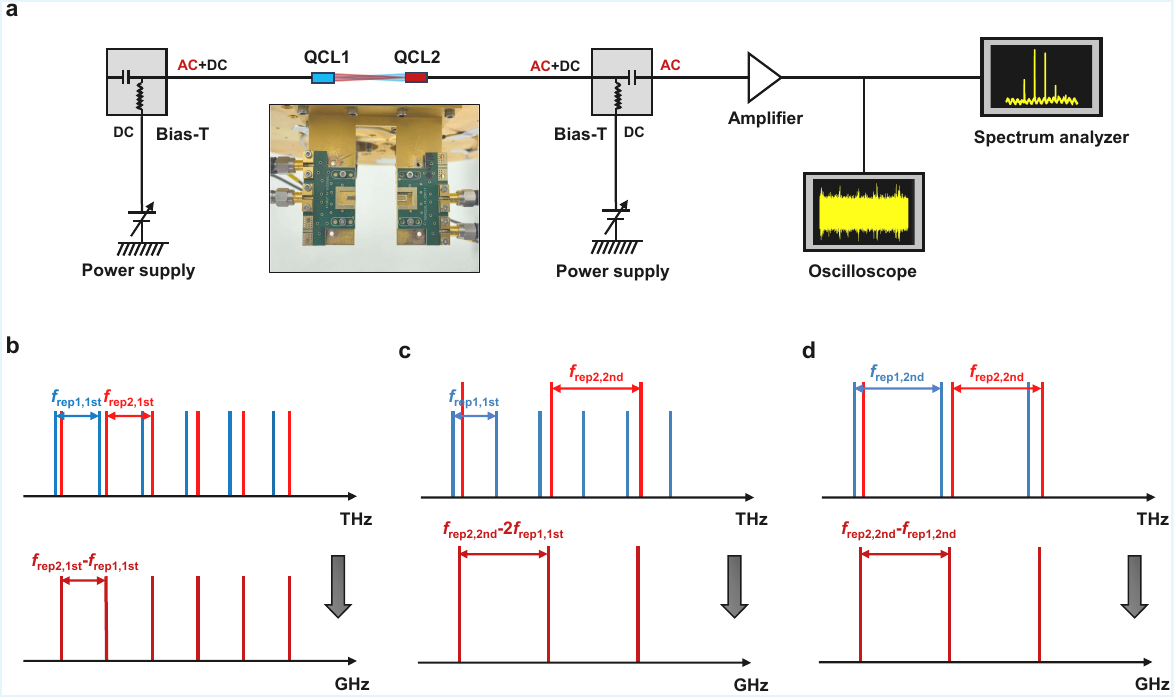}
 \caption{\textbf{a} Experimental setup of the THz dual-comb system. QCL1 and QCL2 are mounted face to face on opposite branches of a Y-shaped cold finger, with a separation of 23~mm between their output facets. The inset shows a photograph of the Y-shaped sample holder integrated with an optimized impedance-matching and high-frequency PCB packaging. \textbf{b--d} Schematic illustration of the reconfigurable dual-comb operation under different comb states. (\textbf{b}) Fundamental–fundamental dual-comb, (\textbf{c}) fundamental–second-harmonic dual-comb, and (\textbf{d})  second-harmonic–second-harmonic dual-comb. In each panel, the upper part shows the spectral distribution of the optical comb lines of the two QCLs in the THz domain, while the lower part depicts the corresponding radio-frequency (RF) spectra obtained via multi-heterodyne down-conversion. Here, $f_{\mathrm{rep1}}$ and $f_{\mathrm{rep2}}$ denote the repetition frequencies of QCL1 and QCL2, respectively, which correspond to either the fundamental ($1\times\mathrm{FSR}$) or harmonic ($n\times\mathrm{FSR}$) comb regimes depending on the operating condition.
 }
 \label{fig1}
\end{figure}

Figure \ref{fig1}(a) shows the experimental setup employed for dual-comb measurements. Both THz QCLs (QCL1 and QCL2) have a nominal cavity length of 4~mm and a ridge width of 150~\textmu m for better comb operation\cite{zhou2019ridge}. The devices are mounted face to face on a Y-shaped cold finger, enabling direct optical coupling without external optical components\cite{li2020toward}. The inset photograph in Figure \ref{fig1}(a) shows the actual packaged device, which is connected to the cryogenic measurement lines via SMA connectors. Owing to the picosecond-scale carrier dynamics of THz QCLs, QCL2 simultaneously functions as both a comb source and a fast detector via the self-detection mechanism\cite{li2022real}.

To enable efficient extraction and transmission of high-frequency signals, including multi-heterodyne and intermode beatnote signals, a dedicated high-frequency printed circuit board (PCB) was designed and wire-bonded to the QCL chips. This optimized packaging minimizes parasitic effects, improves impedance matching, and enhances signal integrity. The high-frequency component of the QCL2 photocurrent is extracted through the AC port of a bias-tee, while the DC bias is supplied through the DC port. The extracted microwave signal is amplified by a 30~dB low-noise amplifier and analyzed using a spectrum analyzer or a high-speed oscilloscope.

Beyond the hardware configuration, the key functionality of the platform lies in its ability to access multiple comb operation regimes within the same cavity architecture. By tuning the driving current and temperature, each QCL can be driven into either a fundamental or a harmonic comb state. This enables the realization of distinct dual-comb configurations, each corresponding to a different repetition-frequency mapping in the multi-heterodyne down-conversion process, as illustrated in Figure \ref{fig1}(b)–(d).

As shown in Figure \ref{fig1}(b), when both QCL1 and QCL2 operate in the fundamental comb regime, their repetition frequencies correspond to the respective FSR, i.e., $f_{\mathrm{rep,1st}} = 1\times \mathrm{FSR}$. The resulting RF dual-comb spectrum exhibits a line spacing of $\Delta f = \lvert f_{\mathrm{rep1,1st}} - f_{\mathrm{rep2,1st}} \rvert$, where $f_{\mathrm{rep1,1st}}$ and $f_{\mathrm{rep2,1st}}$ denote the repetition frequencies of QCL1 and QCL2 in the fundamental comb regime, respectively, corresponding to a fundamental--fundamental dual-comb configuration.

Figure \ref{fig1}(c) illustrates the case where QCL1 operates in the fundamental regime while QCL2 switches to the harmonic comb regime. In the harmonic state, nonlinear intracavity dynamics lead to mode skipping, effectively increasing the repetition frequency to an integer multiple of the FSR. For the second-harmonic regime, $f_{\mathrm{rep,2nd}} = 2\times \mathrm{FSR}$. In this configuration, each comb line of QCL2 beats with a subset of modes of QCL1 determined by the harmonic relation. The resulting RF comb arises from the superposition of multi-heterodyne beat notes generated by multiple pairs of comb lines satisfying the frequency matching condition, ensuring a well-defined mapping between the optical and RF domains: $\Delta f = \lvert 2 f_{\mathrm{rep1,1st}} - f_{\mathrm{rep2,2nd}} \rvert$, corresponding to a fundamental–second-harmonic dual-comb.

As shown in Figure \ref{fig1}(d), when both QCL1 and QCL2 operate in harmonic comb regimes, their repetition frequencies are integer multiples of the cavity free spectral range. Taking the second-harmonic regime as an example, $f_{\mathrm{rep,2nd}} = 2\times \mathrm{FSR}$. The resulting RF dual-comb spectrum has a line spacing of $\Delta f = \lvert f_{\mathrm{rep1,2nd}} - f_{\mathrm{rep2,2nd}} \rvert$, corresponding to a second-harmonic–second-harmonic dual-comb configuration. 

Similar to the previous cases, the RF comb arises from the superposition of multi-heterodyne beat notes generated by multiple pairs of comb lines that satisfy the harmonic frequency-matching condition, preserving a well-defined mapping between optical and RF domains. Compared with the fundamental regime, harmonic comb operation features increased mode spacing and reduced modal density within the same gain bandwidth.

In this framework, the harmonic order effectively provides an additional degree of freedom for tuning the repetition-frequency difference, enabling flexible RF mapping within a single device platform.

\subsection{QCL characterization}

\begin{figure}[!h]
 \centering
 \includegraphics[width=0.9\linewidth]{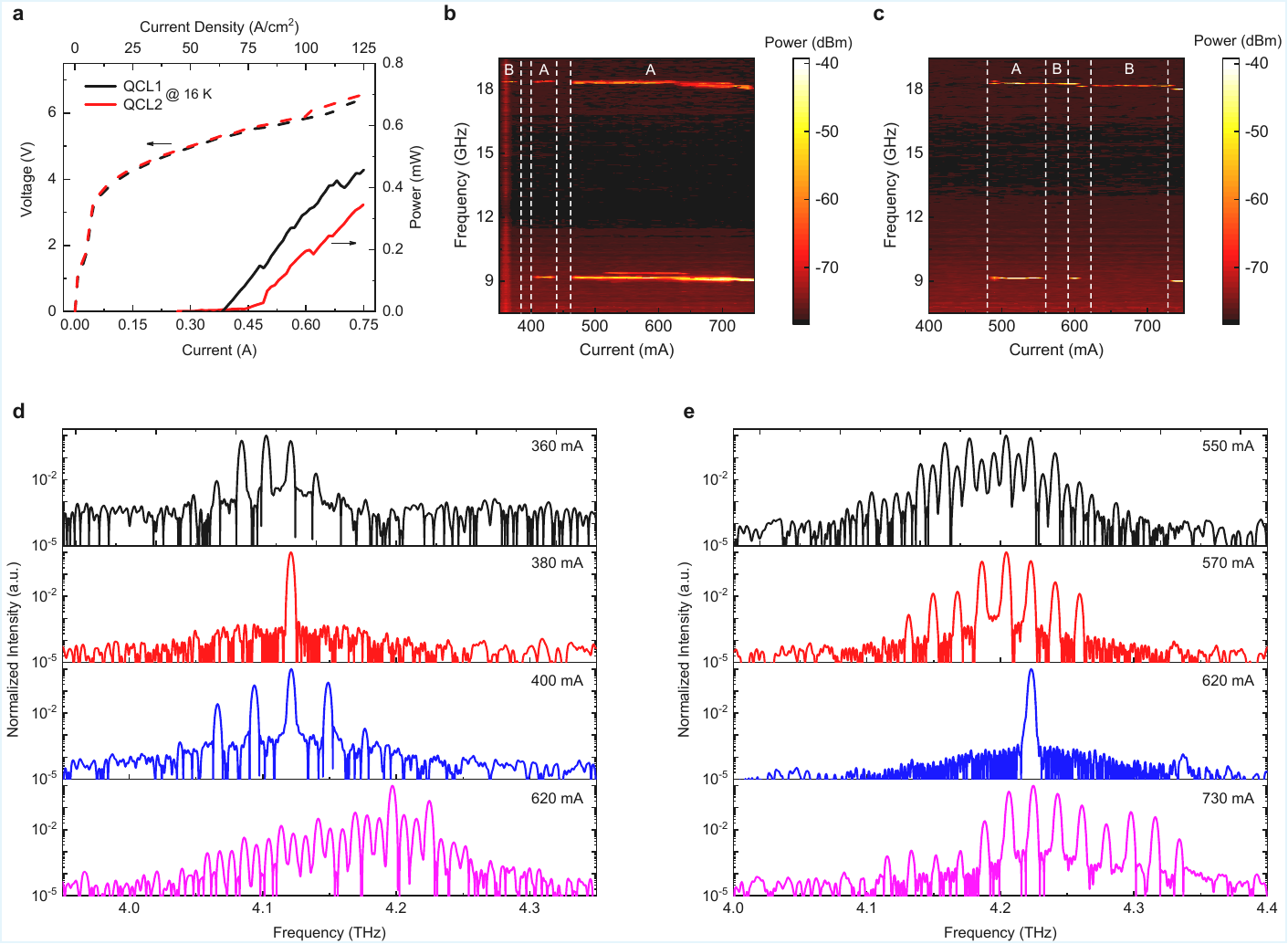}
 \caption{\textbf{a} Light–current–voltage (L–I–V) characteristics of QCL1 and QCL2 measured in continuous-wave (CW) operation at a heat-sink temperature of 16~K. \textbf{b,c} Intermode beatnote spectra of (\textbf{b}) QCL1 and (\textbf{c}) QCL2 under free-running conditions at 16~K, measured with a resolution bandwidth (RBW) of 100~kHz and a video bandwidth (VBW) of 10~kHz. \textbf{d,e} Emission spectra of (\textbf{d}) QCL1 and (\textbf{e}) QCL2 measured using a Fourier-transform infrared (FTIR) spectrometer at different drive currents, illustrating the evolution of comb states under electrical tuning.}

  \label{fig2}
\end{figure}

The performance of the THz QCLs was first characterized under continuous-wave (CW) operation. Figure \ref{fig2}(a) shows the light–current–voltage (L–I–V) characteristics measured at a heat-sink temperature of 16~K. The threshold currents of QCL1 and QCL2 are 380~mA and 450~mA, respectively, with maximum output powers of 0.46~mW and 0.35~mW. The reported power values were measured directly at the cryostat window without correction for optical losses\cite{pistore2024comprehensive}.

To identify the comb operation regimes, intermode beatnote maps were acquired using the self-detection scheme, as shown in Figures~2(b) and 2(c). When a single narrow beatnote appears simultaneously at both $1\times\mathrm{FSR}$ (around 9~GHz) and $2\times\mathrm{FSR}$(around 18~GHz), the QCL operates in the fundamental comb regime. This behavior arises because, in the fundamental comb regime, all adjacent longitudinal modes are phase-locked and contribute to the detected signal. The quadratic detection process generates beatnotes not only between adjacent modes (at $1\times\mathrm{FSR}$), but also between non-adjacent modes (e.g., $2\times\mathrm{FSR}$), leading to the simultaneous presence of multiple harmonic beatnotes. In principle, such beatnotes can extend to higher-order multiples of the FSR\cite{LiOE2022}. This behavior is observed for QCL1 in the current ranges of 400–440~mA and 460–750~mA, and for QCL2 in the range of 480–560~mA (region~A). 

In contrast, when a narrow beatnote is present only at $2\times\mathrm{FSR}$ while the $1\times\mathrm{FSR}$ component disappears, the device enters a harmonic comb regime. This behavior originates from mode skipping in the harmonic comb regime, where only every second longitudinal mode is populated. As a result, no mode pairs exist with a spacing of $1\times\mathrm{FSR}$, and only beatnotes at multiples of the harmonic repetition frequency are observed. This condition is observed for QCL1 in the range of 350–380~mA and for QCL2 in the ranges of 560–590~mA and 620–730~mA (region~B). These results demonstrate that both QCLs can be reversibly switched between fundamental and harmonic comb states by adjusting the driving current.

The comb states are further confirmed by emission spectra measured using a Fourier-transform infrared (FTIR) spectrometer with a spectral resolution of $0.08~\mathrm{cm^{-1}}$, as shown in Figures \ref{fig2} (d) and (e). By tuning the drive current, the devices can transition among single-mode emission, fundamental comb operation, and harmonic comb operation. Notably, QCL1 exhibits a third-harmonic comb regime at specific currents (e.g., 400~mA), corresponding to $f_{\mathrm{rep,3rd}} = 3\times\mathrm{FSR} \approx 27$~GHz. Due to the limited bandwidth of the spectrum analyzer (1~MHz–26.5~GHz), the corresponding beatnote cannot be directly observed in the intermode beatnote map.

The dependence of the comb operation on temperature is provided in Figures~S1 and S2 (Supporting Information). Since temperature strongly affects the operating regime, all dual-comb measurements presented below were performed under stabilized thermal conditions.
Different from the spectrum measurement using the OAP (Off-Axis Parabolic Mirrors) coupling, the dual-comb experiments were carried out using a closed-cycle cryocooler. Although both systems were temperature controlled, differences in thermal environment, sensor position, heat transfer conditions, and the mutual thermal loading resulting from simultaneous operation of two QCLs lead to slight shifts in the operating currents corresponding to individual harmonic states.

\subsection{Dual-comb operation and analysis}

By exploiting the ability to access different harmonic comb states in each QCL, the dual-comb system can be described in terms of a generalized harmonic-order mapping framework. Specifically, the repetition frequencies of the two combs can be expressed as $f_{\mathrm{rep1},\,n_{1}\mathrm{th}} = n_1 \times \mathrm{FSR}_1$ and $f_{\mathrm{rep2},\,n_{2}\mathrm{th}} = n_2 \times \mathrm{FSR}_2$, where $n_1$ and $n_2$ denote the harmonic orders of QCL1 and QCL2, respectively.

Under this framework, the resulting RF dual-comb spectrum is governed not only by the difference between repetition frequencies, but also by the combination of harmonic orders $(n_1, n_2)$, which determines the effective frequency mapping from the THz domain to the RF domain. By tuning $(n_1, n_2)$ through electrical control, distinct dual-comb regimes with different RF line spacing, spectral bandwidth, and mode density can be realized within the same device platform.

In general, the RF frequencies generated by multi-heterodyne mixing can be expressed as
\[
\Delta f = \left| k f_{\mathrm{rep1},\,n_{1}\mathrm{th}} - l f_{\mathrm{rep2},\,n_{2}\mathrm{th}} \right|,
\]
where $k$ and $l$ are integers determined by the harmonic orders and the specific comb-line pairs involved.

\begin{figure}[!h]
 \centering
 \includegraphics[width=0.7\linewidth]{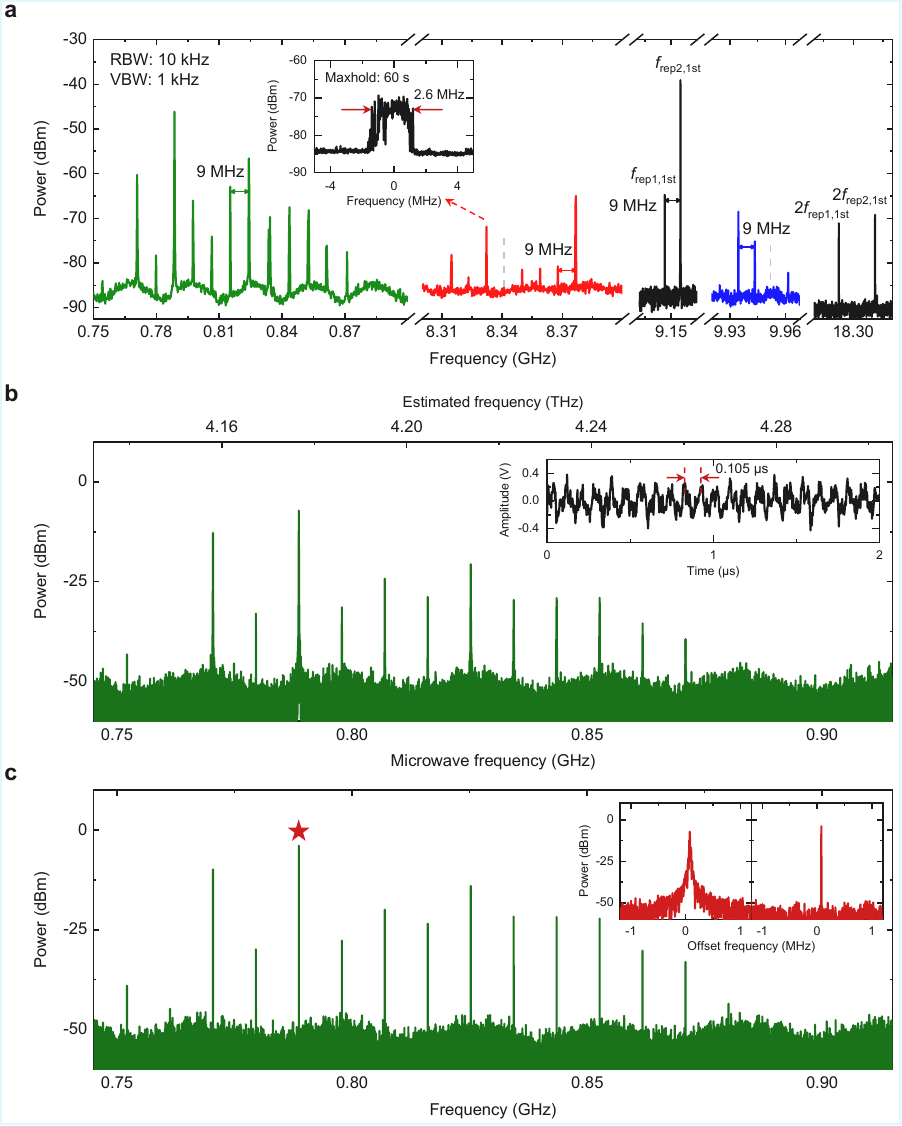}
 \caption{
\textbf{a} Fundamental–fundamental dual-comb signal and corresponding intermode beatnote spectrum measured at a stabilized temperature of 16~K, with an RBW of 10~kHz and a VBW of 1~kHz. The inset shows the maximum-hold (max-hold) spectrum of the mode indicated by the arrow (8.33~GHz), recorded over 60~s, demonstrating the frequency stability of the dual-comb lines. The spacing of the RF comb lines corresponds to the difference in repetition frequencies (\mbox{$\Delta f \approx 9~\mathrm{MHz}$}). \textbf{b} RF dual-comb spectrum obtained via fast Fourier transform (FFT) of the raw time-domain signal. The raw time data corresponds to a single, unaveraged acquisition with a time window of 200~\textmu s (approximately 1800 periods). The inset shows a zoomed-in view of the time-domain signal over a 2~\textmu s interval. The upper horizontal axis indicates the corresponding THz comb frequencies obtained through linear frequency mapping from the RF domain. \textbf{c} RF spectrum after phase correction. The phase correction significantly improves spectral coherence and linewidth performance. The inset compares the selected comb line (marked by the star) before (left) and after (right) correction, with the frequency axis offset such that the line center is set to 0 Hz for clarity. 
}
 \label{fig3}
\end{figure}

\begin{figure}[!t]
 \centering
 \includegraphics[width=0.7\linewidth]{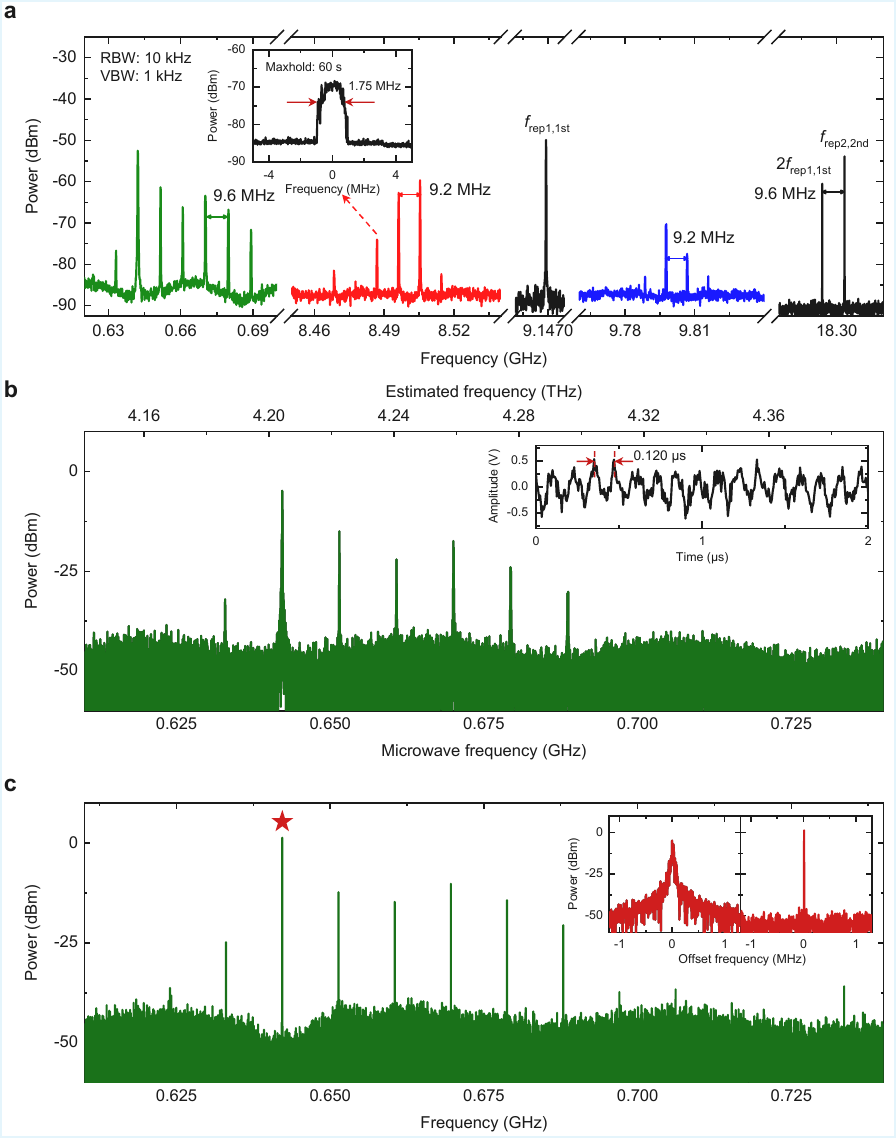}
 \caption{\textbf{a} Fundamental–second-harmonic dual-comb signal and corresponding intermode beatnote spectrum measured at a stabilized temperature of 16~K, with an RBW of 10~kHz and a VBW of 1~kHz. The inset shows the max-hold spectrum of the mode indicated by the arrow (8.487~GHz), recorded over 60~s, exhibiting a linewidth of approximately 1.75 MHz and demonstrating the stability of the comb operation. In this configuration, QCL1 operates in the fundamental comb regime, exhibiting both fundamental and second-harmonic intermode beatnotes, while QCL2 operates in the harmonic comb regime, showing only the second-harmonic beatnote. For the fundamental–second-harmonic configuration, the RF comb spacing is \mbox{$\Delta f \approx 9.6~\mathrm{MHz}$}, as expected from the corresponding repetition-frequency mapping. \textbf{b} RF dual-comb spectrum obtained via fast FFT of the raw time-domain signal. The raw time data corresponds to a single, unaveraged acquisition with a time window of 200~\textmu s (approximately 1900 periods). The inset shows a zoomed-in view of the time-domain signal over a 2~\textmu s interval.  \textbf{c} RF spectrum after phase correction. The recovered optical bandwidth is expanded from 110 GHz to 201 GHz after phase correction. The inset compares the selected comb line (marked by the star) before (left) and after (right) correction, with the frequency axis offset such that the line center is set to 0 Hz for clarity. 
}
 \label{fig4}
\end{figure}

\begin{figure}[!t]
 \centering
 \includegraphics[width=0.7\linewidth]{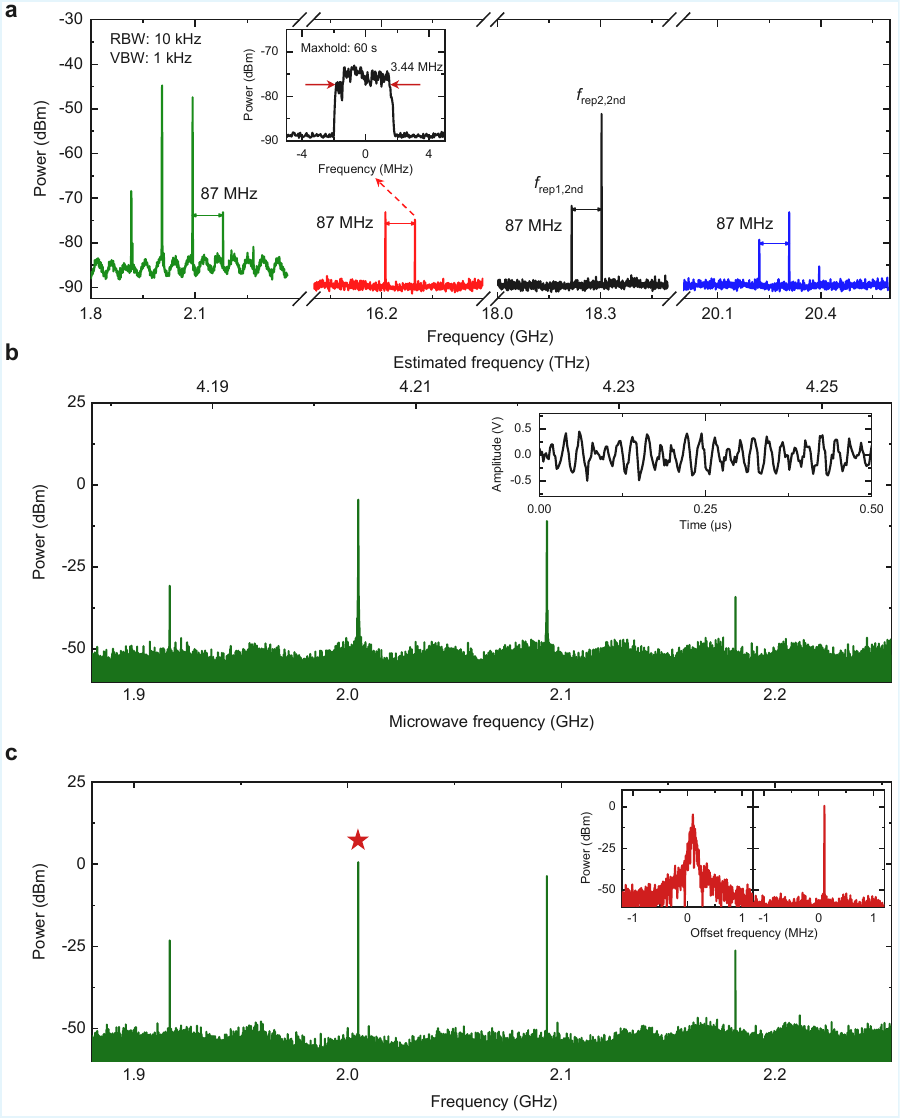}
 \caption{\textbf{a} Second-harmonic–second-harmonic dual-comb signal and corresponding intermode beatnote spectrum measured at a stabilized temperature of 16~K, with an RBW of 10~kHz and a VBW of 1~kHz. The inset shows the max-hold spectrum of the mode indicated by the arrow (16.299~GHz), recorded over 60~s, demonstrating the stability of the harmonic comb operation. In this configuration, both QCL1 and QCL2 operate in the harmonic comb regime, exhibiting only harmonic intermode beatnotes. For the second-harmonic–second-harmonic configuration, the RF comb spacing is \mbox{$\Delta f \approx 87~\mathrm{MHz}$}, as expected from the repetition-frequency difference. \textbf{b} RF dual-comb spectrum obtained via fast FFT of the raw time-domain signal. The trace corresponds to a single, unaveraged acquisition with a time window of 200~\textmu s (approximately 17400 periods). The inset shows a zoomed-in view of the time-domain signal over a 0.5~\textmu s interval.  \textbf{c} RF spectrum after phase correction, showing an increase in the power of the representative comb line marked by the star from $-4.5$~dBm to $0.6$~dBm. The inset compares this comb line before (left) and after (right) correction, with the frequency axis offset such that the line center is set to 0 Hz for clarity. 
}
 \label{fig5}
\end{figure}

By adjusting the driving conditions of the two QCLs, three primary dual-comb configurations—fundamental–fundamental, fundamental–second-harmonic, second-harmonic–second-harmonic—are realized within the same experimental platform shown in Fig. 1, as summarized in Figures \ref{fig3}–\ref{fig5}.

\textbf{Fundamental–fundamental dual-comb.}  

When both QCL1 and QCL2 operate in the fundamental comb regime ($n_1 = n_2 = 1$), the system realizes a conventional dual-comb mapping with RF spacing $\Delta f = |f_{\mathrm{rep1,1st}} - f_{\mathrm{rep2,1st}}|$, serving as a reference case for the harmonic-order framework. This situation corresponds to the schematic shown in Figure \ref{fig1}(b).

When QCL1 and QCL2 were driven at 553~mA and 540~mA, respectively, with the heat sink temperature stabilized at 16~K, a single narrow beatnote appeared at both $1\times\mathrm{FSR}$ and $2\times\mathrm{FSR}$ for each device, as shown in Figure \ref{fig3}(a). The intermode beatnote maps were measured with a resolution bandwidth (RBW) of 10~kHz and a video bandwidth (VBW) of 1~kHz. This confirms that both devices were operating in the fundamental frequency comb regime, and the system entered the fundamental--fundamental dual-comb state. The simultaneous observation of $2\times\mathrm{FSR}$ signal is attributed to the strong optical nonlinearity inherent in THz QCL cavities and the high mutual coherence of the comb modes, which together enable efficient generation of harmonic beat signals at integer multiples of the repetition frequency. Note that due to the limited detection bandwidth in our measurement setup, only the second harmonic ($2\times\mathrm{FSR}$) is presented here. For devices with longer cavities and therefore smaller FSRs, higher-order beatnotes at multiple integer harmonics of the FSR can in principle be observed by the same mechanism. The measured intermode beatnote frequencies for QCL1 ($f_\mathrm{rep1,1st}$) and QCL2 ($f_\mathrm{rep2,1st}$) were 9.146~GHz and 9.155~GHz, with corresponding powers of $-64.74$~dBm and $-39.08$~dBm, respectively. Their frequency difference of approximately $\Delta f \approx 9$~MHz defines the line spacing of the dual-comb spectrum. The lower beatnote power from QCL1 is attributed to the detection scheme: while the signal from QCL2 was directly extracted via self-detection, the signal from QCL1 first underwent free-space coupling to QCL2, incurring additional optical loss.

Figure~\ref{fig3}(a) also presents three sets of dual-comb spectra generated by heterodyne beating between modes of different inter-comb spacings, centered at approximately 0.81~GHz, 8.34~GHz, 
and 9.81~GHz, respectively, all recorded in single-shot mode. In all three cases, the spacing of the RF comb lines is given by the repetition-frequency difference $\Delta f$ determined above. The observation of dual-comb spectra arising from different neighboring comb-line pairs further indicates the good stability and coherence of the dual-comb system. The dual-comb signal in the lowest frequency band (green curve, centered at $\sim$0.81~GHz) exhibits a notably larger number of 
resolved modes and higher signal power compared to the other two sets. This is because this signal originates from multi-heterodyne beating between nearest-neighbor comb modes, where the nonlinear mixing efficiency is highest and the optical spectral overlap is maximal, resulting in the most efficient down-conversion. On both the left (red curve, centered at $\sim$8.34~GHz) and right (blue curve, centered at $\sim$9.81~GHz) sides of the intermode beatnote, individual modes are absent, as marked by the gray dashed lines. This is likely caused by the net-gain distribution and mode-competition effects within the cavity under the given drive conditions, which may 
render certain longitudinal modes weaker than their neighbors due to insufficient gain or strong inter-mode suppression. In addition, the larger frequency separation between non-adjacent 
modes leads to degraded spatial interference overlap and an uneven system frequency response, both of which can reduce the signal-to-noise ratio of the corresponding down-converted dual-comb components below the detection threshold. To assess the long-term stability, the inset shows the ``max-hold'' spectrum of the dual-comb line at 
8.3322~GHz recorded over 60~s, exhibiting a linewidth of approximately 2.6 MHz, which confirms the good mutual coherence of the free-running dual-comb source\cite{10898087}.

To more comprehensively characterize the mutual coherence of the dual-comb system, a high-speed oscilloscope was employed under the experimental configuration shown in Figure~\ref{fig1}(a) to record single-shot (unaveraged) time-domain traces of the 
fundamental--fundamental dual-comb signal. The sampling rate was set to 40~GSa/s to avoid aliasing, and the acquisition window spanned 200~$\mu$s (from $-100~\mu$s to $+100~\mu$s), corresponding to approximately 1800 repetition periods of the dual-comb interference pattern. The inset of Figure~\ref{fig3}(b) shows a magnified view of the time-domain trace over a 2~$\mu$s interval, in which a well-defined periodic structure is clearly visible. The period of this structure is approximately 105 ns, in close agreement with the inverse of the repetition frequency difference $1/\Delta f \approx 111$~ns derived from the spectral measurements. The slight discrepancy is mainly attributed to the free-running operation of the two lasers, for which residual frequency jitter remains in the absence of active locking. This temporal periodicity provides an independent, time-domain confirmation of the mutual coherence between the two combs: only when the relative phase between the two QCL combs remains stable can such a regular interference pattern persist throughout the entire 200~$\mu$s acquisition window. A fast Fourier transform (FFT) was then applied to the full time-domain trace to retrieve the dual-comb spectrum in the microwave domain, as displayed in the main panel of Figure~\ref{fig3}(b). Frequency calibration from the microwave to the terahertz range was performed by matching the highest-power mode in the dual-comb spectrum with the peak emission frequency measured by FTIR under identical drive conditions, and the corresponding terahertz frequencies are indicated on the upper axis.

To further improve the mutual coherence and spectral quality of the dual-comb signal, a computational phase correction based on multiscale extended Kalman filtering (MS-EKF) was applied\cite{burghoff2016computational,burghoff2019generalized,bi2025terahertz}.
This method dynamically tracks and compensates for fluctuations in both the carrier-envelope offset frequency and the repetition frequency difference. The corrected dual-comb spectrum is 
presented in Figure~\ref{fig3}(c). Compared with the uncorrected result, the dual-comb lines exhibit a markedly improved signal-to-noise ratio and noticeably reduced linewidth. Taking 
the representative mode marked by the star symbol as an example, the insets compare the spectral profile before (left) and after (right) correction, with the frequency axis shifted to zero for 
clarity. After correction, the mode power increased from $-7.2$~dBm to $-3.9$~dBm. Moreover, several modes that were previously buried in the noise floor became resolvable after correction, including the spectral lines near 0.76~GHz and 0.88~GHz, which extended the recovered optical bandwidth from 101~GHz to 128~GHz. This bandwidth can serve as a reference for comparison with the other dual-comb configurations.

\textbf{Fundamental--second-harmonic dual-comb.}

When QCL1 operates in the fundamental regime ($n_1 = 1$) and QCL2 in a second-harmonic regime ($n_2 = 2$), the dual-comb system enters a cross-harmonic mapping configuration. In this case, the RF comb spacing is no longer governed by a simple difference between repetition frequencies, but instead arises from a mixed harmonic relation, $\Delta f = \lvert 2 f_{\mathrm{rep1,1st}} - f_{\mathrm{rep2,2nd}} \rvert$, reflecting the coupling between different harmonic orders. This situation corresponds to the schematic shown in Figure \ref{fig1}(c).

When QCL1 and QCL2 were driven at 565~mA and 540~mA, respectively, with the heat sink temperature stabilized at 16~K, the intermode beatnote map reveals that QCL1 exhibits beatnotes at both 
$1\times\mathrm{FSR}$ and $2\times\mathrm{FSR}$, while QCL2 shows only a single beatnote at $2\times\mathrm{FSR}$, as shown in Figure~\ref{fig4}(a). This confirms that QCL1 operates in the fundamental comb regime and QCL2 in the harmonic comb regime, indicating that the system has switched into the fundamental–second-harmonic dual-comb state. The intermode beatnote frequencies are 9.1469~GHz for QCL1 and 18.303~GHz for QCL2, with corresponding powers of $-49.9$~dBm and $-53.9$~dBm, respectively. In this configuration, the dual-comb line spacing is no longer determined simply by the difference between two FSR frequencies, but rather by the frequency offset between the harmonic beatnote of QCL2 and twice the fundamental beatnote of QCL1, i.e., $\Delta f = f_\mathrm{rep2,2nd} - 2f_\mathrm{rep1,1st} = 9.6\ \mathrm{MHz}$. This modified frequency mapping demonstrates that by tuning the drive current to switch one device between the fundamental and harmonic comb regimes, the effective repetition frequency difference, which determines the RF comb spacing, can be actively controlled. Such tunability offers a flexible means to adjust the spectral sampling density and acquisition rate of the dual-comb system.

Figure~\ref{fig4}(a) also presents three sets of dual-comb spectra centered at approximately 0.66~GHz, 8.49~GHz, and 9.80~GHz, recorded in single-shot mode with an RBW of 10~kHz and a VBW of 1~kHz. 
Since these three sets were not acquired simultaneously, their comb line spacings differ slightly: 9.6~MHz for the lowest-frequency set and 9.2~MHz for the other two, which reflects the small frequency drift of the free-running system between measurements. As in the fundamental--fundamental case, the lowest-frequency set (green curve) exhibits the highest mode count and signal power, consistent with the higher nonlinear mixing efficiency between nearest-neighbor modes. The inset shows the ``max-hold'' spectrum of the dual-comb mode at 8.487~GHz recorded over 60~s, exhibiting a linewidth of approximately 1.75~MHz. This linewidth, which is comparable to that in the fundamental–fundamental case, confirms that coherent dual-comb operation is maintained even when one device operates in the harmonic comb regime.

Following the same acquisition procedure described above, a single-shot time-domain trace was recorded over a 200~$\mu$s window (approximately 1900 repetition periods) and Fourier-transformed to yield the RF dual-comb spectrum shown in Figure~\ref{fig4}(b). The inset shows a zoomed-in view of the 
time-domain signal over a 2~$\mu$s interval, in which a periodic interference pattern with a period of $\sim$120~ns is observed, in reasonable agreement with $1/\Delta f \approx 109$~ns derived from the 
9.2~MHz comb spacing. The upper horizontal axis indicates the corresponding THz frequencies obtained via linear frequency mapping from the RF domain.

After applying the MS-EKF phase correction, the dual-comb spectrum (Figure~\ref{fig4}(c)) exhibits significantly enhanced signal-to-noise ratio and narrowed linewidth. For the representative mode marked by the star, the power increased from $-4.8$~dBm to $1.3$~dBm after correction. Notably, the 
recovered optical bandwidth broadened from 110~GHz to 201~GHz, representing a nearly twofold improvement compared with the uncorrected result and surpassing the bandwidth achieved in the 
fundamental--fundamental case. This bandwidth expansion is consistent with the broader gain profile accessible when one device operates in the harmonic comb regime, as evidenced by the emission spectra shown in Figures~\ref{fig2}(d) and 2(e).

\textbf{Second-harmonic–second-harmonic dual-comb.}  

When both QCLs operate in the second-harmonic comb regime ($n_1 = n_2 = 2$), the system follows a harmonic–harmonic mapping configuration, in which the effective repetition frequencies are increased and the RF comb spacing reflects the difference between higher-order harmonic beatnotes.

When the drive currents of QCL1 and QCL2 were further adjusted to 720~mA and 565~mA, respectively, intermode beatnotes from both devices were observed only at $2\times\mathrm{FSR}$, while the beatnotes at $1\times\mathrm{FSR}$ disappeared completely. This indicates that both lasers were operating in the harmonic comb regime. Consequently, the dual-comb signal switched from the fundamental--second-harmonic to the second-harmonic--second-harmonic dual-comb state. The corresponding intermode beatnote frequencies, $f_{\mathrm{rep1},\,2\mathrm{nd}}$ and $f_{\mathrm{rep2},\,2\mathrm{nd}}$, were measured to be 18.216 GHz and 18.303 GHz, with powers of $-71.7$~dBm and $-51.1$~dBm, respectively. The resulting dual-comb mode spacing, given by the difference between these two frequencies, is 87~MHz as shown in Figure~\ref{fig5}(a).

Figure~\ref{fig5}(a) also presents three sets of dual-comb spectra centered at approximately 2.0~GHz, 16.3~GHz, and 20.2~GHz, respectively, all recorded with the same RBW and VBW setting. Compared with the 
fundamental--fundamental and fundamental--second-harmonic cases, the number of observable comb lines in each set is significantly reduced. Specifically, the lowest-frequency set (green curve) contains approximately 4 modes, the middle set (red curve) approximately 2 modes, and the highest-frequency set (blue curve) approximately 3 modes. This reduced mode count is a direct consequence of the harmonic comb regime itself: when both QCLs operate at $2\times\mathrm{FSR}$, the effective mode spacing is doubled and the total number of modes within the gain bandwidth is approximately halved, leaving fewer lines available for multi-heterodyne beating. In addition, the relatively large difference between the drive currents shifts the spectral centers of the two QCLs apart, reducing their spectral overlap and further limiting the number of observable beatnotes. The inset shows the 
``max-hold'' spectrum of the dual-comb mode at 16.299~GHz recorded over 60~s, exhibiting a linewidth of approximately 3.44 MHz. Notably, this linewidth is comparable to those measured in the fundamental--fundamental and fundamental--second-harmonic cases, indicating that the long-term frequency stability of the dual-comb signal is maintained regardless of the harmonic order. Phase noise spectra for all three representative modes were also measured; the results are presented in Figure~S3 (Supporting Information), further indicating comparable stability across the different dual-comb configurations.

Following the procedure described above, a single-shot time-domain trace was recorded over a 200~$\mu$s window, corresponding to approximately 17400 repetition periods given the larger $\Delta f \approx 87$~MHz. The inset shows a zoomed-in view over a 0.5 µs interval, in which a well-defined periodic structure is clearly visible. This time-domain periodicity again independently confirms the mutual coherence of the two harmonic combs. After Fourier transformation, the resulting RF dual-comb spectrum (Figure~\ref{fig5}(b)) shows a reduced number of comb lines compared with the previous 
two configurations, which is a direct consequence of the intrinsically lower mode density in the harmonic comb regime.

As in the previous cases, MS-EKF phase correction was applied to the dual-comb signal. 
For the representative mode marked by the star, the power increased from $-4.5$~dBm to $0.6$~dBm. 
However, owing to the intrinsically limited mode density of the harmonic comb regime, the phase correction did not yield a further increase in the number of detectable comb lines, in contrast to the bandwidth expansion observed in the fundamental--fundamental and fundamental–second-harmonic states.

\textbf{Higher-order harmonic dual-comb (second-harmonic–third-harmonic).}

To further generalize this framework, we investigate dual-comb operation between higher-order harmonic states with unequal harmonic orders ($n_1 = 3$, $n_2 = 2$). This configuration represents a generalized harmonic-order mapping regime, where multiple multi-heterodyne beatnotes arise from different combinations of harmonic modes, leading to a more complex but predictable RF spectral structure.

\begin{figure}[!h]
 \centering
 \includegraphics[width=0.8\linewidth]{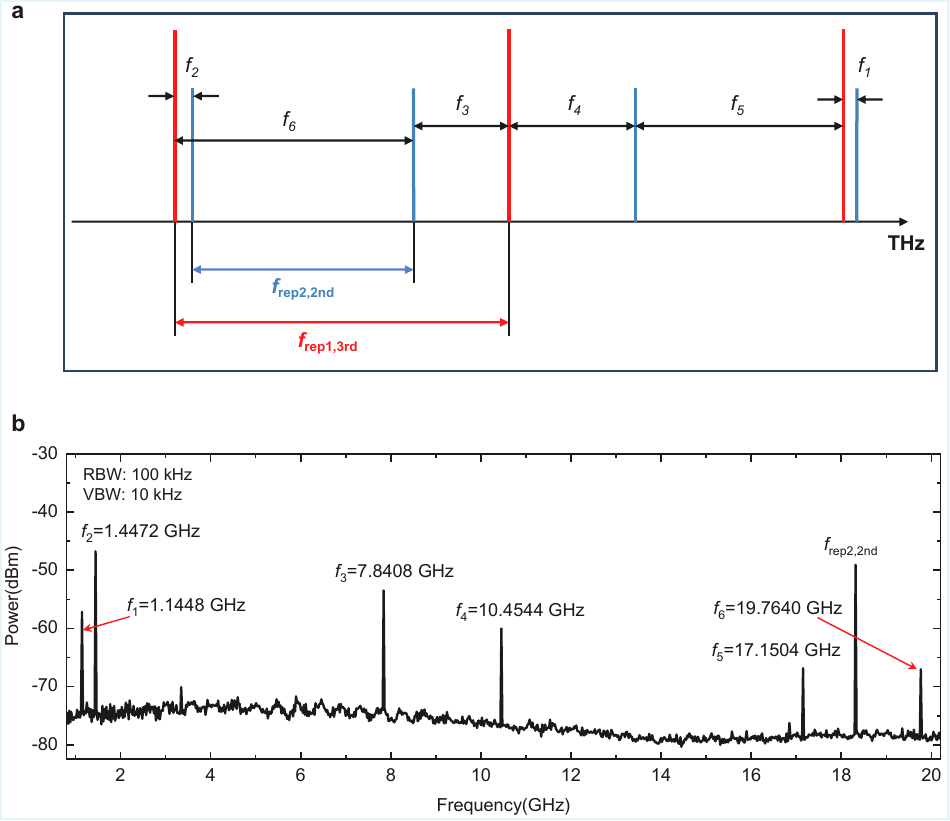}
 \caption{\textbf{a} Schematic illustration of the beat-frequency mapping between a second-harmonic comb (QCL2) and a third-harmonic comb (QCL1). Here, $f_{\mathrm{rep1}}$ and $f_{\mathrm{rep2}}$ denote the repetition frequencies of QCL1 and QCL2, respectively. The diagram shows the expected multi-heterodyne beatnote positions arising from higher-order harmonic mixing. \textbf{b} Second-harmonic–third-harmonic dual-comb signal and corresponding intermode beatnote spectrum measured at a stabilized temperature of 16~K, with an RBW of 100~kHz and a VBW of 10~kHz. In addition to the repetition frequency of QCL2 ($f_{\mathrm{rep2,2nd}}$), multiple beatnotes labeled $f_1$ to $f_6$ are observed, corresponding well to the predicted positions shown in \textbf{a}. The agreement between experiment and schematic confirms the controllable multi-heterodyne mapping across different harmonic orders.
}
 \label{fig6}
\end{figure}

Building on the observation that QCL1 enters the third-harmonic comb regime at elevated drive currents (Figure~\ref{fig2}(d)), we further explore multi-heterodyne mixing between higher-order harmonic states. Owing to the different thermal environments of the FTIR characterization platform and the closed-cycle dual-comb system, slight shifts in the operating current corresponding to each harmonic state are expected, while the harmonic-state assignment is verified by the deterministic multi-heterodyne frequency relations described below. Figure~\ref{fig6}(a) schematically illustrates the beat-frequency mapping when QCL1 operates in the third-harmonic regime ($3\times\mathrm{FSR}$, red) and QCL2 operates in the second-harmonic regime ($2\times\mathrm{FSR}$, blue). 

Within the spectral overlap region, each mode of QCL1 beats with its nearest mode of QCL2, producing a series of beatnote signals labeled in ascending frequency order as $f_1$ to $f_6$, and they satisfy the following relations:

\begin{equation}
f_3 + f_6 = f_4 + f_5 = f_{\mathrm{rep1,3rd}}
\tag{1}
\end{equation}
\begin{equation}
f_1 + f_5 = f_3 + f_4 = f_6 - f_2 = f_{\mathrm{rep2,2nd}}
\tag{2}
\end{equation}

where $f_\mathrm{rep,3rd}$ and $f_\mathrm{rep,2nd}$ denote the effective repetition frequencies of the third-harmonic and second-harmonic combs, respectively. Beatnotes arising from more widely spaced mode pairs are not labeled due to the limited detection bandwidth of the spectrum analyzer. These relations represent a direct experimental manifestation of the generalized harmonic-order mapping, confirming that the RF spectrum is governed by deterministic combinations of comb-line indices across different harmonic orders.

To verify the theoretical prediction, the drive currents of QCL1 and QCL2 were set to 565~mA and 620~mA, respectively. The resulting RF spectrum and intermode beatnotes, measured with an RBW of 100~kHz and a VBW of 10~kHz, are displayed in Figure~\ref{fig6}(b). Multiple beatnote signals ($f_1$ to $f_6$) are clearly observed at the expected frequency positions. The extracted frequencies satisfy the constraint relations given by Eq. (1) and (2), confirming that these signals indeed originate from controlled multi-heterodyne mixing between the second-harmonic and third-harmonic comb states. Together with the previously characterized harmonic-comb operating regions shown in Figure 2, these deterministic frequency relations provide the basis for the harmonic-state assignment under the dual-comb measurement conditions. From this correspondence, the repetition frequency of QCL1 in the third-harmonic comb regime is inferred to be approximately 27.6~GHz and the repetition frequency of QCL2 in the second-harmonic comb regime is measured to be approximately 18.5~GHz. Interestingly, the observed RF comb does not follow the simple difference between the fundamental repetition frequencies. Instead, it arises from a near-commensurate relation between higher-order harmonics of the two combs. In this case, the repetition frequencies satisfy $2f_{\mathrm{rep1,3rd}} \approx 3f_{\mathrm{rep2,2nd}}$, leading to an effective RF comb spacing given by
\[
\Delta f_{\mathrm{eff}} = |2f_{\mathrm{rep1,3rd}} - 3f_{\mathrm{rep2,2nd}}| \approx 300~\mathrm{MHz}.
\]

This reduced spacing originates from multi-heterodyne mixing between comb-line pairs whose indices satisfy the approximate integer relation $(k,l) = (2,3)$, resulting in a fine RF comb structure with significantly smaller spacing than the fundamental repetition frequencies. 

Among the resolved signals, $f_1$ and $f_2$ correspond to the lowest-frequency dual-comb components arising from the second--third-harmonic mixing. Although only a limited number of comb lines were resolved, constrained by the finite number of accessible harmonic modes and the reduced heterodyne mixing efficiency between modes of different harmonic orders, the observation of $f_1$ to $f_6$ with the correct frequency relations provides compelling evidence that higher-order harmonic dual-comb operation is feasible and can be controlled through drive-current tuning. 

However, the fundamental--third-harmonic configuration was not investigated. This is attributed to the shift in the absolute emission spectrum of the QCLs induced by drive-current tuning: when QCL1 operates in the third-harmonic regime and QCL2 in the fundamental regime, the spectral overlap between the two lasers becomes insufficient to generate detectable beatnote signals.

\subsection{Discussion}

Beyond the demonstration of reconfigurable dual-comb operation, several experimental observations provide further insights into the system performance and potential optimization strategies.

First, the linewidth of individual RF comb lines exhibits a dependence on the comb configuration. In the fundamental regime, the linewidth is measured to be on the order of a few megahertz (2.6 MHz), while in the fundamental–second-harmonic configuration a slightly narrower linewidth (1.75 MHz) is observed. In contrast, the second-harmonic–second-harmonic configuration shows a broader linewidth (3.44 MHz). Although the underlying mechanism requires further investigation, these results suggest that the comb dynamics and coherence properties are influenced by the harmonic order and the corresponding mode distribution. Phase noise measurements, however, do not reveal a significant difference among these regimes within the current experimental resolution. This also suggests that flexible switching between different dual-comb states can be achieved without introducing appreciable additional phase noise.

Second, the application of phase-correction techniques proves effective across all comb states, including harmonic regimes. As shown in Figures S4-S6 (Supporting Information), the fluctuations of both $\Delta f_{\mathrm{rep}}$ and $\Delta f_{\mathrm{ceo}}$ are effectively tracked and compensated, further confirming the validity of the phase-correction approach. This enables improved spectral coherence and facilitates the extraction of weak signals in the multi-heterodyne spectrum. Such capability is particularly relevant for extending the detection sensitivity and may support future implementations of real-time phase correction and adaptive signal processing.

From a system perspective, the demonstrated control over harmonic states provides a new route for spectral manipulation in the THz domain. In contrast to conventional approaches relying on external optical components, the repetition-frequency mapping can be engineered directly through the intrinsic nonlinear dynamics of the QCL. This may enable simplified implementations of frequency-to-space mapping. Specifically, when combined with dispersive elements such as blazed gratings, the harmonic comb states with comb spacing several-fold larger than that of the fundamental one can facilitate line-by-line spectral selection and parallel spatial channelization. Moreover, the self-detection capability of the QCL platform enables coherent detection via a reflective collection geometry, suggesting that the demonstrated harmonic dual-comb architecture could be extended to applications such as high-resolution spectral imaging and multi-channel spatial mapping. In addition, the present results suggest that, through further device optimization, harmonic dual-comb operation with a larger number of resolvable comb lines may be achieved by maintaining better spectral alignment between the two QCLs while accessing different harmonic states. Such improvement would further expand the application potential of harmonic dual-combs in spectroscopy, frequency metrology, and multi-channel THz systems. Moreover, stabilization strategies that have proven effective for fundamental QCL combs are also likely to be extendable to harmonic comb operation.

It is also worth noting that, although the present platform does not provide continuously tunable repetition rates within a single comb state, the results here show that harmonic order itself can serve as an additional degree of freedom for dual-comb engineering\cite{senica2026continuously}. By switching between fundamental and different harmonic comb states, the effective repetition-frequency mapping can be reconfigured within the same THz QCL platform, leading to different RF line spacings, mode densities, and multi-heterodyne pathways without modifying the cavity structure or relying on external optical components. In this sense, while the tunability enabled here is discrete rather than continuous, it still substantially expands the operational flexibility of the system and provides a complementary route toward enhanced controllability in compact semiconductor dual-comb sources.

Looking forward, further improvements may be achieved by integrating spectral filtering or self-referencing schemes, such as isolating a single comb line for feedback or injection, to enhance comb stability and signal-to-noise ratio. In addition, preliminary observations indicate that harmonic comb states can also be accessed under pulsed operation, suggesting the possibility of synchronized dual-comb systems with improved temporal control.

While the detailed microscopic mechanisms underlying harmonic comb formation including gain competition, spatial hole burning, gain defects, effective Rabi oscillations and more remain subjects of active investigation, the present work focuses on the system-level functionality enabled by harmonic-order switching. Rather than providing a comprehensive physical model, this study demonstrates that harmonic order can be reliably controlled and exploited as a practical engineering degree of freedom for reconfigurable dual-comb operation. The robust multi-heterodyne signals observed across different harmonic configurations further verify the stability and practicality of this approach. The recent theoretical work by Silvestri et al. further complements the current physical understanding by identifying the resonance between the effective Rabi frequency and a cavity mode as a key mechanism for self-starting harmonic comb formation in QCLs\cite{carlo2026Rabi}. Their findings suggest that the harmonic order may be intrinsically linked to the cavity mode structure and the gain medium's ultrafast dynamics, which is consistent with our observation that harmonic states can be reliably accessed and switched via electrical and thermal control.

\section{Conclusion}

In summary, we have demonstrated a reconfigurable multi-harmonic dual-comb system based on a self-detected THz QCL platform. By precisely controlling the driving current and thermal conditions, multiple dual-comb configurations—including fundamental–fundamental, fundamental–second-harmonic, second-harmonic–second-harmonic, and higher-order harmonic (second-harmonic–third-harmonic) dual-combs—have been realized within the same device architecture. 

These results establish harmonic order as an additional degree of freedom for dual-comb operation, enabling flexible and controllable multi-heterodyne mapping without modifying the cavity structure or introducing external optical components. The experimental observations further confirm that stable dual-comb operation can be achieved across different comb regimes, including asymmetric configurations where the two QCLs operate in distinct harmonic states. This work provides a compact and versatile approach for THz dual-comb generation and frequency mapping, offering new opportunities for high-resolution spectroscopy, frequency metrology, and integrated THz systems. Potential applications include high-resolution molecular spectroscopy, rapid pharmaceutical quality control, and THz wireless communication systems requiring multiple carrier frequencies.


\section{Experimental Section}

\threesubsection{THz QCLs and characterizations}

The THz QCLs employed in this work are based on a GaAs/AlGaAs hybrid active region incorporating bound-to-continuum transitions for THz photon emission and resonant-phonon depopulation of the lower laser state. Detailed layer thickness and doping profiles are described in ref.~\cite{WanSR}. The entire active region was grown on a semi-insulating GaAs(100) substrate by molecular beam epitaxy (MBE) and subsequently processed into single plasmon waveguide laser ridges with a ridge width of 150 $\mu$m. Laser ridges with a cavity length of 4 mm were cleaved, corresponding to a free spectral range (FSR) of approximately 9.15 GHz (${\rm FSR}=c/2nL$, where $n\approx3.6$ is the effective refractive index and $L=4$ mm), and then indium-bonded onto copper heat sinks for wire bonding.

For electrical and optical characterizations, the QCL devices were mounted onto the cold plane of a closed-cycle cooler (Acryo-4K-MB) operating at temperatures down to 6 K. The average optical power was measured using a calibrated thermal power meter (Ophir, 3A-P THz) positioned in front of the vacuum chamber window, with no additional optical coupling elements. The driving currents of both QCLs were precisely controlled by high-stability current sources (Qube CL, ppq sense), synchronized with the power meter through an automation platform to enable high-precision power characterization. The intermode beatnote of each QCL comb and the multi-heterodyne dual-comb signals were measured via self-detection using one of the QCLs as the ultrafast THz detector (carrier relaxation time in the picosecond range) \cite{LiOE}. A custom-designed printed circuit board (PCB) was employed for electrode contact and RF signal extraction. The PCB incorporates impedance-matching microstrip lines optimized for 50 $\Omega$ systems, effectively bridging the $\sim$20 $\Omega$ impedance of the THz QCL to the characteristic impedance of the measurement instrumentation. This design ensures efficient transmission of both intermode beatnote signals and multi-heterodyne dual-comb signals. RF spectra were recorded using a signal and spectrum analyzer (Rohde \& Schwarz, FSW26), while phase noise spectra were measured using a phase noise analyzer (Rohde \& Schwarz, FSWP26). Time-domain waveforms of the dual-comb interferograms were acquired using a high-speed real-time oscilloscope (Teledyne Lecroy, WaveMaster 820 Zi-B) operating at a sampling rate of 40 GSa/s.

\threesubsection{Extended Kalman filtering for phase correction}

To mitigate the detrimental effects of phase noise on dual-comb coherence, a computational phase-correction algorithm based on a multiscale Extended Kalman Filter (EKF) framework was implemented. In a THz QCL dual-comb system, mutual coherence fluctuations arise primarily from two sources: the carrier-envelope offset frequency difference ($\Delta f_{\rm ceo}$) and the time-varying repetition-rate difference ($\Delta f_{\rm rep}$). The multiscale EKF operates directly on the time-domain interferograms acquired by the high-speed oscilloscope and simultaneously estimates and corrects both parameters by recursively predicting and updating the instantaneous values of $\Delta f_{\rm rep}(t)$ and $\Delta f_{\rm ceo}(t)$ \cite{burghoff2019generalized}. In our previous work \cite{bi2025terahertz}, this approach was successfully demonstrated in a THz QCL dual-comb system with high repetition-rate detuning, where it effectively restored mutual coherence and achieved uncertainty-limited linewidths across the full detection bandwidth. The corrected time traces are subsequently processed by Fast Fourier Transform (FFT) to obtain high-resolution RF dual-comb spectra.

\medskip
\textbf{Supporting Information} \par 
Supporting Information is available from the Wiley Online Library or from the author.

\medskip
\textbf{Acknowledgements} \par 
This work is supported by the National Key Research and Development Program of China (2025YFE0217500), the National Science Fund for Distinguished Young Scholars (62325509), the National Natural Science Foundation of China (62505344, 62235019, 62531005, 62575299, 62275258, 62305364, 62435017, and T2550072), the Scientific Instrument and Equipment Development Project of the Chinese Academy of Sciences (PTYQ2026YZ0050), the CAS Project for Young Scientists in Basic Research (YSBR-069), and Autonomous deployment project of State Key Laboratory of Materials for Integrated Circuits (SKLJC-Z2025-B03), Science and Technology Commission of Shanghai Municipality (23ZR1474000). The authors gratefully acknowledge the technical and equip-ment support provided by ShanghaiTech Material and Device Lab (SMDL).

\medskip

%
\bibliographystyle{MSP}
\bibliography{REF.bib}




\end{document}